\documentclass{svjour2}

\def\C{\mathbb C}

\def\E{\mathbb E}

\def\P{\mathbb P}
\def\R{\mathbb R}

\def\Z{\mathbb Z}

\def\ffi{{\varphi}}

\def\wh{\widehat}

\def\cS{{\mathcal S}}

\def\dist{{\rm{dist}}}

\def\supp{{\rm{supp}}}
\def\one{{\mathbf 1}}

\def\u0{{\mathbf 0}}

\def\uu{{\mathbf u}}

\def\ux{{\mathbf x}}

\def\uH{{\mathbf H}}
\def\uDelta{{\mathbf \Delta}}
\def\uU{{\mathbf U}}
\def\uV{{\mathbf V}}
\def\wh{\widehat}

\def\eps{{\epsilon}}
\def\Om{{\Omega}}
\def\om{{\omega}}
\def\Lam{{\Lambda}}
\def\Gam{{\Gamma}}

\def\rs{\rm s}

\def\rv{\rm v}
\def\rV{\rm V}

\def\rx{\rm x}
\def\ry{\rm y}
\def\rz{\rm z}

\def\pmn{\par\medskip\noindent}
\def\psn{\par\smallskip\noindent}

\def\z2{{\Z^2}}
\def\zp2{{\Z^2_{\geq}}}

\def\dist{{\,{\rm dist}}}

\def\Bmf{\mathfrak B}

\def\truc#1#2#3{\smash{\mathop{\,\, #1 \,\, }\limits^{#2}_{#3}}}

\def\bfw{\mathbf w}

\def\bfV{\mathbf V}

\def\BGam{{\mbox{\boldmath${\Gamma}$}}}
\def\BLam{{\mbox{\boldmath${\Lam}$}}}

\usepackage{graphics}

\usepackage{latexsym}
\usepackage{amsfonts, amssymb}
\usepackage{amsmath}
\usepackage{array}
\begin{document}
\title{ Wegner-type bounds for a two-particle\\ continuous
Anderson model\\ with an alloy-type external potential}

\author{A. Boutet de Monvel$^1$ \and V. Chulaevsky$^2$
\and P. Stollmann$^3$\and Y. Suhov$^4$}
%
\institute{
{\scriptsize{
$^1$Institut de Math\'{e}matiques de Jussieu\\
Universit\'{e} Paris 7\\
175 rue du Chevaleret, 75013 Paris, France\\
E-mail: aboutet@math.jussieu.fr
\pmn
$^2$D\'{e}partement de Math\'{e}matiques \\
Universit\'{e} de Reims, Moulin de la Housse, B.P. 1039, \\
51687 Reims Cedex 2, France \\
E-mail: victor.tchoulaevski@univ-reims.fr
\pmn
$^3$ Fakult\"{a}t f\"{u}r Mathematik\\
Technische Universit\"{a}t Chemnitz\\
09107 Chemnitz, Germany\\
E-mail: peter.stollmann@mathematik.tu-chemnitz.de
\pmn
$^4$ Statistical Laboratory, DPMMS\\
University of Cambridge, Wilberforce Road, \\
Cambidge CB3 0WB, UK\\
E-mail: Y.M.Suhov@statslab.cam.ac.uk
}}
}

\date{}

%
%
\maketitle
\pmn
%
\begin{abstract}
We consider a  two-particle quantum systems in $\R^d$ with
interaction and in presence of a random external potential (a continuous
two-particle Anderson model). We establish Wegner-type estimates
(inequalities)
for such models, assessing the probability that random spectra of
Hamiltonians in finite volumes intersect a given set.
\end{abstract}

\section{Introduction. The two-particle\\ Hamiltonian in the continuum}\label{intro}
\pmn

This paper is concerned with a two-particle Anderson model in $\R^d$ with
interaction. The Hamiltonian $\uH\left(=\uH(\omega )\right)$ is a Schr\"{o}dinger operator
of the form $-\frac{1}{2}\uDelta+\uU(\ux)+\uV(\omega;\ux)$ acting on functions
$\phi\in L_2(\R^{2\cdot d})$. This means that we consider two particles, each living in $\R^d$, in the following fashion: $\ux=(x_1, x_2)$ and each $x_j=\big({\rx}_j^{(1)},\ldots,{\rx}_j^{(d)}\big)$
represents the coordinates of the $j$'s particle. Here, $-\frac{1}{2}\uDelta$ is the standard kinetic energy resulting from adding up the kinetic energies $-\frac{1}{2}\Delta_j$ of the different particles and assuming that we are dealing with particles of identical masses. In case of different masses,  $-\frac{1}{2}\uDelta$ would have to be replaced by $\uH_0=-\frac{1}{2}\sum_{j=1, 2}\frac{1}{m_j}\Delta_j$, without changing any of the analysis involved. The potential $\uU(\cdot )$ is, as usually, identified with the corresponding multiplication operator. It incorporates the interaction between particles, as well as possibly a deterministic external potential.
We assume

(D) \emph{Boundedness of the interaction potential:}
  $$\begin{array}{cl} \uU\in L^\infty(\R^{2 d})
\end{array}\eqno (1.1)$$
All the particles are subject to the same random external random potential
$V(x;\omega)$ where $\omega$ runs through a probability space $\Omega$. The respective potential energy appearing in the Hamiltonian $\uH$ results from adding the potential energies of the single particle and thus reads
$$\begin{array}{cl} \uV(\ux;\omega)=\sum_{j=1, 2}V(x_j;\omega)
\end{array}\eqno (1.2)$$
In this paper we consider an alloy-type random potential $V(z;\omega)$ with specific properties listed in the next section. A note on the notation: we usually denote two-particle quantities by boldface letters.

As is known in the one-particle case (see \cite{CL,PF,St2} and the references therein) and in the discrete two-particle case, \cite{CS2}, the disorder introduced through the random field can generate a pure point spectrum, i.e.,   \emph{Anderson localisation}. In the present paper we take the first step in showing Anderson localisation in the continuum two-particle case as well by proving Wegner bounds (we defer the multi-scale analysis of this case to a future work, \cite{BCSS}).

These bounds assess concentration of the eigenvalues of $\uH_\BLam(=\uH_{\BLam}(\om ))$,
a finite-volume version of Hamiltonian $\uH$. Operator $\uH_\BLam$ acts vectors in $L_2(\BLam )$:
$$
\begin{array}{cl}\uH_{\BLam}(\om )=-\frac{1}{2}\uDelta+\uU(\ux)+\uV(\omega;\ux)\\
\mbox{  with Dirichlet boundary conditions at  }\partial\BLam\\
\end{array}
\eqno (1.3)
$$
Here $\BLam\subset\R^{2d}$ is a rectangle which we call a two-particle box, or simply a box, of the form
$$
\BLam=\Lam^{(1)}\times \Lam^{(2)},
\eqno(1.4)
$$
where $\Lam^{(j)}\subset\R^d$ is a cube with edges parallel to the
coordinate axes in $\R^d$, $j=1, 2$. To be more specific, given
$L_1,L_2>0$ and $u_1, u_2\in\R^d$, we set henceforth:
$$
\BLam_{L_1,L_2}(\uu )=\Lam_{L_1}(u_1)\times \Lam_{L_2}(u_2),
\eqno (1.5)
$$
where $\uu =(u_1, u_2)\in\R^{2 d}$ and, for $L>0$ and
$v=\big({\rv}^{(1)},\ldots ,{\rv}^{(d)}\big)$,
$$\Lam_L(v)=
{\operatornamewithlimits{\times}\limits_{i=1}^d}
\big[-L+{\rv}^{(i)},{\rv}^{(i)}+L\big].
\eqno (1.6)
$$

We would like to note that the methods adopted in this paper
are applicable to a wide range of other boundary conditions
including periodic and `elastic' (e.g., Neumann b.c.). In fact, what we
need is that operator $\uH_\BLam$ in (1.4) is self-adjoint and
has a compact resolvent $\left(\uH_\BLam -{\rz}I\right)^{-1}$ for
nonreal ${\rz}\in\C\setminus\R$; this covers all `classical'
Krein's self-adjoint extensions, from the `soft' (Neumann b.c.)
to `hard' (Dirichlet b.c.).

In what follows, the short notation $\BLam$ is used for a two-particle
box $\BLam_{L_1, L_2}(\uu )$, when the parameters $L_1$, $L_2$ and $\uu$ are
unambiguous. Similarly, $\BLam'$ is a shorthand notation for
$\BLam_{L'_1, L'_2}(\uu')$.

Under the conditions imposed in this paper (see Eqns (2.2)--(2.8)),
operator $\uH_\BLam$ has a compact resolvent and therefore a discrete spectrum consisting of eigenvalues of finite multiplicity. It is convenient to write these eigenvalues
$E^{({\BLam})}=E^{({\BLam})}(\om)$ in increasing order:
$$
E^{({\BLam})}_0\leq E^{({\BLam})}_1\leq E^{({\BLam})}_2\leq\ldots .
\eqno (1.7)
$$
The `one-volume' Wegner bound assesses the probability
$$
\P\left(\exists\;\hbox{ $k$
with $\left|E-E^{(\BLam )}_k\right|$}\leq\eps\right),\eqno (1.8)$$
that at least one eigenvalue $E^{(\BLam )}_k$ of operator $H_\BLam$
falls in a (narrow) interval around a given point $E$ on the
spectral axis. The `two-volume' Wegner bound deals with
$$
\P\left(\exists\;\hbox{ $k$ and $k'$
with $E^{(\BLam )}_k,E^{(\BLam')}_{k'}\in I$
and $\left|E^{(\BLam )}_k-E^{(\BLam' )}_{k'}\right|$}\leq\eps\right).
\eqno (1.9)
$$
This is the probability that some eigenvalues $E^{(\BLam )}_k$
and $E^{(\BLam')}_{k'}$ of the operators  $\uH_{\BLam}$
and $\uH_{\BLam'}$ come close to each other in a given interval
$I\subset\R$ of the spectral axis, for two (distant)
two-particle boxes $\BLam$ and  $\BLam'$. Here and below,
$\P$ stands for the corresponding probability measure on the
underlying probability space (specified below).
\pmn

{\bf Remark.} From the probabilistic
point of view, bounds (1.8) and (1.9) are examples of concentration inequalities,
albeit for rather implicit RVs $E^{(\BLam )}_k$ and
$E^{(\BLam')}_{k'}$ carrying a considerable amount of dependence. For a single-particle
Anderson model, under natural assumptions
on the character of the random terms in Hamiltonians (1.3),
the Wegner bounds are rather straightforward. We do not provide here an extensive
bibliography on this subject; apart from the original work by Wegner, \cite{W}, we refer to the references in the monographs,
\cite{CL,PF,St2},  the surveys \cite{KM,V} as well as in a recent paper by Combes et al. \cite{CHK}.
For a two-particle continuum systems, these estimates have not been studied
before. (A version of the Wegner bounds for the so-called
tight-binding two-particle model (a discrete modification of
the model treated here) was established in \cite{CS1}; discrete multi-particle Wegner-type bounds can also be found in \cite{AW,K})
\pmn

In the next section, we give formal conditions upon the structure
of the potential energy term in Eqn (1.2).
In what follows, $\|\cdot\|_{\max}$ denotes the sup-norm in
$\R^{2d}$.

Throughout the paper, $|\BLam |$ stands for the (Euclidean) volume
of a set $\BLam\subset\R^{2 d}$ and $|\Lam |$ for that
of a set $\Lam\subset\R^d$. We also use the similar notation
$|\BGam |$ and $|\Gam |$ for the cardinality of lattice subsets
$\BGam\subset\Z^{2 d}$ and $\Gam\subset\Z^d$; in particular,
$\Pi_j\BGam$ stands for the cardinality of the projection $\Pi_j\BGam$,
$j=1.2$. Here
$$
\Z^d=\big\{s=\big({\rs}^{(1)},\ldots,{\rs}^{(d)}\big):
\;{\rs}^{(i)}\in\Z,\;\;i=1,\ldots ,d\big\}\eqno (1.10)
$$
is the integer lattice canonically embedded in $\R^d$.
\psn

\section{External random potentials of alloy-type}
\label{extptntl}
\pmn

In this paper, the random external potential
$V(x;\omega )$, $x\in\R^d$, $\om\in\Omega$, is assumed to be of alloy-type, over a cubic lattice:
$$
V(x;\om )=\sum_{s\in\Z^d}{\rV}_s(\om)\varphi_s(x-s).
\eqno (2.1)
$$
Here $\mathbf{V}=({\rV}_s,\;s\in\Z^d)$, is a family of real
random variables (RVs) ${\rV}_s$ on some probability space $(\Om,{\mathfrak B},\P )$ and $\{\varphi_s,\;s\in\Z^d\}$ is a (nonrandom)
collection of `bump' functions $y\in\R^d\mapsto\varphi_s(y)$.
In probabilistic terms, $\mathbf{V}$ is a real-valued random field
(RF) on $\Z^d$. Physically speaking, the RV ${\rV}_s$
represents the amplitude of an `impurity' at site $s$ of lattice $\Z^d$
while the function $\varphi_s$ describes the `propagation' of
the impact of this impurity across the space $\R^d$.

To avoid excessive technicalities concerning self-adjointness of our Hamiltonians $\uH_{\BLam}$ we assume that the alloy type random potential is uniformly bounded via:

(E0) \emph{Boundedness of the random field}:
$$
\sup_{s\in\Z^d}\|{\rV}_s\|_\infty =:M<\infty\eqno (2.2)
$$

(E1) \emph{Boundedness of the bump functions}: $\ffi_s$ are bounded non-negative functions, with
$$
\truc{\sup}{}{x\in\R^d}
\left[ \sum\limits_{s\in\Z^d}\;\varphi_s(x-s)\right]<+\infty,
\;\;\forall\;x\in\R^d .\eqno (2.3)
$$

We will also need a lower bound:

(E2) \emph{Covering condition}:
$$
\sum\limits_{s\in\Lam_L(u)\cap\Z^d}\;\varphi_s(x-s)
\geq\;1,\;\;\forall\;L\geq 1,\;u\in\R^d,\;x\in\Lam_L(u).
\eqno (2.4)
$$

We stress the fact that we do not need independence of the random variables ${\rV}_s$ for different sites. What we need is a regularity requirement for the induced conditional marginal distribution.


Given a site $s\in\Z^d$, consider the conditional distribution function
$$F\left(\,{\ry} \big| \Bmf_{\{s\}^{\rm c}}\right):=
\P\big({\rV}_s<{\ry}\big|\Bmf_{\{s\}^{\rm c}}\big),\eqno (2.5)
$$
relative to the sigma-algebra $\Bmf_{\{s\}^{\rm c}}$ generated by RVs $V_t,\;t\in\Z^d\setminus\{s\}$.
Next, set:
$$
\begin{array}{l}\nu (\eps )
:= {\operatornamewithlimits{\sup}\limits_{s\in\Z^d}}\;\;
{\operatornamewithlimits{\sup}\limits_{\ry\in \R}}\;\;
{\operatornamewithlimits{{\rm{sup\,ess}}}\limits_{
{\bfV}_{\{s\}^{\rm c}}}}
\;\Big[F\left(\,{\ry}+\eps  \big| \Bmf_{\{s\}^{\rm c}}\right)-
F\left(\,{\ry} \big| \Bmf_{\{s\}^{\rm c}}\right)\Big]\,.\end{array}
\eqno (2.6)
$$

The following condition is general enough so as to cover a large class of external
random potentials, e.g., regular Gaussian random fields as well as some Gibbsian random fields.
Notice, however, that it can be further relaxed. In this paper, we do not seek maximal generality,
preferring a maximal simplicity of presentation.

(E3) \emph{Uniform marginal control}: the marginal probability distributions of ${\rV}_s$, conditional on $\bfV_{\{s\}^{\rm c}}$, admit uniformly bounded probability density functions (PDF)
$$
p_s({\ry};\Bmf_{\{s\}^{\rm c}}) = {\rm d} F\left(\,{\ry} \,\big|\, \Bmf_{\{s\}^{\rm c}}\right)/{\rm d}{\ry}
$$
such that, for all $\eps\in(0,1)$,
$$
\rho_\infty := {\operatornamewithlimits{\sup}\limits_{s\in\Z^d}}\;\;
{\operatornamewithlimits{\sup}\limits_{\ry\in \R}}\;\;
{\operatornamewithlimits{{\rm{sup\,ess}}}\limits_{
{\Bmf}_{\{s\}^{\rm c}}}}
\;p_s({\ry};\Bmf_{\{s\}^{\rm c}}) < \infty.
\eqno (2.7)
$$
As a consequence, $\nu(\eps) \leq \rho_\infty\eps$.

(E4) \emph{Finite propagation range:} functions $\varphi_s$
have a bounded support: $\exists$ $R\in (0,\infty )$ with
$$
\varphi_s(y)=0\;\hbox{ whenever }\;||y||_{\max}>R.
\eqno(2.8)
$$

\pmn

{\bf Remarks.} (1) The condition (1.1) as well as (E0), (E1) and (E3) are stronger than what is needed. In particular, it is easy to see that it suffices to require H\"{o}lder continuity of the conditional cumulative distribution function (CDF):
$F\left(\,{\ry} \,\big|\, \Bmf_{\{s\}^{\rm c}}\right)$, uniform with respect to the condition:
$$
\begin{array}{l}{\operatornamewithlimits{\sup}\limits_{s\in\Z^d}}\;\;
{\operatornamewithlimits{\sup}\limits_{\ry\in \R}}\;\;
{\operatornamewithlimits{{\rm{sup\,ess}}}\limits_{
{\bfV}_{\{s\}^{\rm c}}}}
\;\Big[F\left(\,{\ry}+\eps  \big| \Bmf_{\{s\}^{\rm c}}\right)-
F\left(\,{\ry} \big| \Bmf_{\{s\}^{\rm c}}\right)\Big]\,
\le \eps^b,
\end{array}
\eqno (2.7')
$$
for all $\eps\in[0,1]$ and some $b\in(0,1)$. (Moreover, one can assume log-H\"{o}lder continuity of the conditional CDF
with appropriately chosen constants.) However, the above condition (2.7) is quite popular and gives rise to slightly simpler notations.

(2) Condition (E3) and the quantity $\nu (\eps )$ can easily be understood in the independent case. Denote by $\mu_s=\P\circ \rV_s$ the law of ${\rV}_s$ which is a measure on the real line and by $s(\mu_s;\eps)$ its modulus of continuity, i.e., $s(\mu_s;\eps)=\sup_{a\in\R}\mu([a,a+\eps])$. The latter is always bounded by $1$ and so is
$$
\nu (\eps )=\sup_{s\in\Z^d}s(\mu_s;\eps)
$$
in this particular case.

(3) There are interesting correlated ensembles for which condition (E3) is well established, see \cite{C1}.

(4) The regularity condition imposed in \cite{AW} (\textbf{Assumption R}) also yields $\nu (\eps )\le \rho_\infty\cdot\eps$, in our notation.

(5) For measure-theoretic concepts used above, see \cite{D}, Appendix; \cite{E}, Theorem 3.1.; \cite{F}, Chapter V; \cite{GS}, Chapter 4.

\section{A one-volume Wegner-type bound}
\label{1vlbnd}

The one-volume Wegner-type bound for two-particle finite-box Hamiltonians
$\uH_{\BLam}$ is given in Theorem \ref{ThmOneVol} below.
Let $\Sigma\left(\uH_{\BLam}\right)$ denote the (random) spectrum
of (random) operator $\uH_{\BLam}$, i.e., the countable set
$$\Sigma\left(\uH_{\BLam}(\om)\right)
=\{
E^{(\BLam)}_k(\om): \; k=0, 1, \ldots \}
$$
of its eigenvalues.

\begin{theorem}\label{ThmOneVol}
 Assume that \emph{(D)} and \emph{(E0-E3)} are satisfied. Then there is a constant $C$ such that for all boxes $\BLam=\Lam^{(1)}\times \Lam^{(2)}$, all $E\in\R$ and $\eps\in (0,1)$:
$$\P\Big(\big[E,E+\eps\big]\cap
\Sigma\left(H_{\BLam}\right)\neq\emptyset\Big)
\le C(1+E\vee 0)^{\frac{d}{2}}| \BLam |\cdot \min_j| \Lam^{(j)}|\cdot\nu(\eps) .
\eqno (3.1)$$
\end{theorem}

The expression in the RHS of (3.1) includes the
`volume' factors $| \BLam |$ and $\min_j| \Lam^{j}|$ with different meaning.
The first one, together with $C(1+E\vee 0)^{\frac{d}{2}}$, comes from
an upper bound of the number of eigenvalues of $\uH_{\BLam}$ below $E+1$.
The second one, together with $\nu(\eps)$, comes from the concentration bound for each individual eigenvalue based on \cite{C1,St1,St2} which we summarize in Lemma \ref{LemmaSt} below.

We will need the following

\begin{definition}\label{DefDM}
Consider a Euclidean space $\R^q$ and its positive orthant $\R^q_+$.
A function $\Phi:\, \R^q \to\R$ is called diagonally-monotone (DM) if
\psn
{\rm (i)} $\Phi({\bfw}+{\mathbf r})\geq \Phi({\mathbf v})$
$\forall$ ${\mathbf r}\in\R^q_+$ and any ${\bfw}\in \R^q$,
\psn
{\rm (ii)} $\Phi({\bfw} + t{\mathbf e}) - \Phi({\bfw})
\geq t$, $\forall$ ${\bfw}\in\R^q$ and $t>0$,
where ${\mathbf e}= (1, 1, \ldots, 1)\in \R^q$.
\end{definition}

\begin{lemma}\label{LemmaSt}
Let $J$ be a finite set with $|J|\geq 2$, and $\mu$ be a probability measure on $\R^J$.
For every $j\in J$, denote by $\mu_j( \cdot \; ;x_{J\setminus\{ j\}})$ the marginal probability measure
induced by $\mu$ on the $j$-th coordinate ${\rx}^{(j)}$ conditional on $x_{J\setminus\{ j\}}$.
Assume that $\mu_j$ admits a uniformly bounded PDF $p_j( {\ry};x_{J\setminus\{ j\}})$,
$\ry \in \R$:
$$
c(\mu) := \truc{\sup {\rm ess}}{ }{x_{J\setminus\{ j\}} \in \R^{J\setminus\{ j\}}} \|p_j( \;\cdot\;;x_{J\setminus\{ j\}})\|_\infty < \infty.
$$
\pmn
Further, let  $\Phi:\R^J\to\R$ be a DM function. Then we have the following concentration bound, for any $a\in\R$:
$$
\mu\{x\in\R^J:\,\Phi(x)\in [a,a+\eps]\}\le | J| \cdot c(\mu) \cdot\eps.
$$
\end{lemma}
\proof. See \cite{C1}, Lemma 4.2, where this result is proved in a more general context.

\qed

\pmn
{\it Proof of Theorem \ref{ThmOneVol}.}

Our assumptions on the potential term imply that $\uU(\cdot)+\uV(\,\cdot \,;\om)$
is uniformly bounded. By the Weyl's formula (\cite{RS}) we
know that $E^{(\BLam)}_k(\om)\ge E+1$ for $k\ge C_1(1+E\vee 0)^{\frac{d}{2}}$ where $C_1>0$ is a constant. Therefore,
$$
\P\Big(\big[E,E+\eps\big]\cap
\Sigma\left(H_{\BLam}\right)\neq\emptyset\Big)\le\sum_{k\le C(1+E\vee 0)^{\frac{d}{2}}}
 \P\Big( E^{(\BLam)}_k\in \big[E,E+\eps\big]\Big) .
$$
We will now proceed to prove that every term in the above sum can be estimated by
$\min_j| \Lam^{(j)}|\cdot\nu(\eps)$, showing the desired bound. Fix $k$ and chose $j_0\in\{1, 2\}$
such that $| \Lam^{(j_0)}|=\min\;[| \Lam^{(1)}|, | \Lam^{(2)}|]$. Next, set $J= \Lam^{(j_0)}\cap \Z^d$.
We will show that the conditional probability for
the $k$th eigenvalue to fall in $[E,E+\eps]$ is bounded:
$$
\truc{\rm sup\; ess}{}{\Bmf_{\Z^d\setminus J}}
\P \big( E^{(\BLam)}_k \in \big[E,E+\eps\big]| \Bmf_{\Z^d\setminus J} \big)\le |J|\cdot\nu(\eps).
$$
\psn
Here $ \Bmf_{\Z^d\setminus J}$ stands for the sigma-subalgebra
of $\Bmf$ generated by ${\mathbf V}_{\Z^d \setminus J} =\{ {\rV}_s, \;s\in \Z^d\setminus J \}$.

We now aim to use the concentration bound from Lemma \ref{LemmaSt} above, for a fixed realisation
${\mathbf V}_{\Z^d \setminus J}$. Here
$\mu$ is identified with $\P_J(\;\cdot\;|{\mathbf V}_{\Z^d \setminus J})$, the restriction of the conditional
 distribution $\P(\;\cdot\;|\Bmf_{\Z^d \setminus J})$ to the 'complementary' sigma-algebra $\Bmf_{J}$ generated
 by ${\mathbf V}_J = \{ {\rV}_{s}: \, s\in J \}$, conditional on ${\mathbf V}_{\Z^d \setminus J}$.  Then the quantity
 $c(\mu;\eps)$ is bounded by $\nu(\eps)$ (this follows from the  definition of $\nu(\eps)$ and assumption (E3). Furthermore, $|J|=|\Lam^{(j_0)}|=\min_{j=1,2} | \Lam^{(j)}|$, by our agreement. Setting $\Phi_k({\mathbf V}_J):= E^{(\BLam)}_k$, it remains to prove that $\Phi_k$ is a DM function. Recall, we are working with eigenvalues of the operator
$$
\uH_\BLam = -\frac{1}{2}\uDelta+\uU(\ux)+\uV(\omega;\ux),\mbox{  on  }L^2(\BLam)
$$
where
$$
\uV(\omega;\ux)=\sum_{j=1}^2\sum_{s\in\Z^d}\om_s\cdot \varphi_s(x_j-s).
$$
Since all the bump functions are nonnegative, operator $\uH_\BLam$ is DM in the variables
$V_s, \, s\in J$. Now $\Phi_k({\mathbf V}_J+t\cdot {\mathbf  e}_J)$ is the $k$-th eigenvalue of
$$\begin{array}{l}
-\frac{1}{2}\uDelta+\uU+\sum\limits_{j=1}^2\sum\limits_{s\in S} {\rV}_s\cdot \varphi_s(x_j-s)\\
\qquad \qquad
+\sum\limits_{j=1}^2\sum\limits_{s\in J}({\rV}_s+t)\cdot \varphi_s(x_j-s) \\
\qquad \qquad\qquad \qquad = \uH_\BLam(\om)+\left[\sum\limits_{j=1}^2 \sum\limits_{s\in J}t\cdot \varphi_s(x_j-s)\right],\end{array}$$
where the function in square brackets acts as multiplication. By assumption (E2), we know that
the corresponding multiplication operator is bounded below by $t{\mathbf I}$, where ${\mathbf I}$ stands for the identity operator. Then the min-max principle
 for the eigenvalues gives that $\Phi_k({\mathbf v`V}_J+t\cdot
 {\mathbf e}_J)\ge \Phi_k({\mathbf V}_J)+t$. Therefore, $\Phi_k$ is a DM function.
Thus, Lemma \ref{LemmaSt} applies and we get the desired bound (3.1).
\qed
\pmn

{\bf Remarks.} (1) Our $2$-particle Wegner-type bound has precursors concerning discrete
Schr\"{o}dinger operators, see \cite{C1,CS1} and \cite{K}, where some particular boxes have been treated
explicitly.

(2) The recent work \cite{AW} addresses Wegner-type bounds for correlated potentials in the discrete
setting. There the focus is on joint distributions of eigenvalues, not on multi-particle models.

\section{A two-volume Wegner-type bound}
\label{2vlbnd}

In this section we state and prove a two-volume
Wegner-type bound; see Theorem \ref{ThmTwoVol} below.
As was mentioned earlier, the two-volume Wegner-type bound
(cf. Eqn (1.9)) is established for a pair of
boxes $\BLam=\BLam_{L_1,L_2}(\uu )$ and $\BLam'=\BLam_{L'_1,L'_2}(\uu')$
(more precisely, for the spectra $\Sigma\left(H_{\BLam}\right)$ and
$\Sigma\left(H_{\BLam'}\right)$ of the corresponding Hamiltonians
$H_{\BLam}$ and $H_{\BLam'}$),
under an assumption that the distance between
$\BLam$ and $\BLam'$ is of the
same order of magnitude as the size of these boxes. Such bounds are an important ingredient in the variable energy multi-scale analysis based on \cite{vDK}. See \cite{St2} and \cite{CS2} for a discrete two-particle version. Due to the dependence inherent in the two-particle case, the multi-scale analysis has to be changed sbstantially; we defer this to a future publication \cite{BCSS}.

Given $L_1,L_2,L'_1,L'_2>1$ and $\uu,\uu'\in\R^d\times\R^d$, we call boxes $\BLam=\BLam_{L_1,L_2}(\uu )$ and $\BLam'=\BLam_{L'_1,L'_2}(\uu')$
\emph{sufficiently distant}, if
$$\begin{array}{l}
\min\, \{\|\uu -\uu'\|_{\max}, \|\cS(\uu) -\uu'\|_{\max} \}\\ \qquad\qquad
>  8\max\{L_1+R,L_2+R,L'_1+R,L'_2+R\},
\end{array}
\eqno(4.1)
$$
where $\cS(\uu)$ denotes the reflected point $\cS(\uu)=(u_2,u_1)$ and $R$ is the constant from (E4).
A useful notion is the \emph{shadow} $\Pi\BLam$ of a two-particle box
$\BLam$:
$$\Pi\BLam =\Pi_1\BLam\cup\Pi_2\BLam .
\eqno (4.2)
$$
As before, $\Pi_1\BLam$ denotes the projection of $\BLam$
to the first and $\Pi_2\BLam$ the projection to the second
Cartesian factor $\R^d$ in $\R^d\times\R^d$.

\begin{lemma}\label{SepLem}
\label{L4.1} Consider two boxes
$\BLam=\BLam_{L_1,L_2}(\uu )$ and $\BLam'=\BLam_{L'_1,L'_2}(\uu')$ that are sufficiently distant and define $\wh{\BLam}=\BLam_{L_1+R,L_2+R}(\uu )$ and
$\wh{\BLam}'=\BLam_{L'_1+R,L'_2+R}(\uu')$. Then at least one of
the following five possibilities will occur:
$$
\begin{array}{lll}
{\rm{(A)}} \qquad{} & \Pi_1\wh{\BLam}\cap\left[\Pi_2\wh{\BLam}\cup
\Pi\wh{\BLam}'\right]&=\emptyset,\\ \;\\
{\rm{(B)}}\qquad{} &  \Pi_2\wh{\BLam} \cap\left[\Pi_1\wh{\BLam} \cup
\Pi\wh{\BLam}'\right]&=\emptyset,\\ \;\\
{\rm{(C)}} \qquad{} & \Pi_1\wh{\BLam}'\cap\left[\Pi\wh{\BLam} \cup
\Pi_2\wh{\BLam}'\right]&=\emptyset,\\ \;\\
{\rm{(D)}} \qquad{} & \Pi_2\wh{\BLam}'\cap\left[\Pi\wh{\BLam} \cup
\Pi_1\wh{\BLam}'\right]&=\emptyset,\\ \;\\
{\rm{(E)}} \qquad{} &  \Pi\wh{\BLam} \cap \Pi\wh{\BLam}'=\emptyset.
\end{array}
$$
\end{lemma}
\begin{proof}
 See \cite{CS1}, Lemma 2.1.
\end{proof}

In the case (E) above, we will say that the boxes $\wh{\BLam}$ and $\wh{\BLam}'$ are
completely separated, while in cases (A)--(D) they will be called partially separated. Note that the partial separation is not incompatible with the complete one.

\begin{corollary}\label{CorSep}
 Consider two boxes
$\BLam=\BLam_{L_1,L_2}(\uu )$ and $\BLam'=\BLam_{L'_1,L'_2}(\uu')$ that are sufficiently distant. Then after renaming
the boxes and/or re-ordering the coordinate projections, one of the following two possibilities will occur:
$$\begin{array}{ll}{\rm{(I)}}
& (V_s;s\in\Pi \BLam\cap\Z^d) \mbox{ is independent of }(V_s;s\in\Pi \BLam'\cap\Z^d)\\
{\rm{(II)}}& (V_s;s\in\Pi_1\BLam\cap\Z^d) \mbox{ is independent of }
(V_s;s\in (\Pi_2\BLam \cup \Pi \BLam') \cap\Z^d).\\
\end{array}$$
\end{corollary}
\begin{proof}
 Using Lemma \ref{SepLem} and renaming, if necessary, the boxes or their projections, we may assume that (A) or (E) of  Lemma \ref{SepLem} occurs. Consider the case (A), for definiteness. From the definition of the external random fields, Eqns (1.2) and (2.1), we get that every
$\uV(\ux;\om)\one_{\BLam'}(x)$ depends only on those $V_s$, for which $\supp\, \varphi_s(\cdot-s)$ intersects either $\Pi_1\BLam'$ or $\Pi_2\BLam'$. By (E4) this requires that
$\Lam_R(s)\cap \left[\Pi_1\BLam'\cup \Pi_2\BLam'\right]\neq\emptyset$. By (A) this is not the case for $s\in\Pi_1\BLam\cap\Z^d$, as claimed. \qed
\end{proof}

\begin{theorem}\label{ThmTwoVol}
Assume that \emph{(D)} and \emph{(E0-E4)} are satisfied. Then, for every interval $I=[a,b]\subset\R$, there exists
a constant $C_2(I)>0$ such that for every pair
$\BLam$ and $\BLam'$ of sufficiently distant boxes and  every $\eps\in(0,1)$ we have that
$$\begin{array}{l}
\P\big(\dist\left[\Sigma\left(H_{\BLam}\right)\cap I,
\Sigma\left(H_{\BLam'}\right)\cap I\right]
\leq \eps \big)\\
\qquad\leq C_2(I)\,\cdot\,
\left|\BLam\right|\;\left|\BLam'\right|\;{\operatornamewithlimits{\max}\limits_{j=1,2}}\;
\max\;\Big[\big|\Pi_j\BLam\big|,\big|\Pi_j\BLam'\big|\Big]
\,  \nu(2\eps) .\end{array}
\eqno(4.3)
$$
\end{theorem}

\proof
 We assume that we are in case (I) of Corollary \ref{CorSep} (complete separation). Set $J=\Pi_1\BLam$. As in the proof of Theorem 1 we estimate the probability in question by conditioning
 on $\Bmf_{\Z^d \setminus J}$:
$$\begin{array}{l}
\P\big(\dist\left[\Sigma\left(H_{\BLam}\right)\cap I,
\Sigma\left(H_{\BLam'}\right)\cap I\right]
\leq \eps \big)\\
=\E\big(\P\big(\dist\left[\Sigma\left(H_{\BLam}\right)\cap I,
\Sigma\left(H_{\BLam'}\right)\cap I\right]
\leq \eps| \Bmf_{\Z^d \setminus J} \big)\big).
\end{array}
\eqno(4.4)
$$

We now estimate the inner probability: first, there are at most $C_3(R,I)| \BLam|$, respectively,  $C_3(R,I)| \BLam'|$ eigenvalues of $H_{\BLam}$ and $H_{\BLam'}$ in the interval $I$. Moreover, by the above Corollary \ref{CorSep}, $H_{\BLam'}$ and consequently its eigenvalues are independent of ${\mathbf V}_J$, so we label them
$E^{(\BLam' )}_{k'}({\mathbf V}_{\Z^d \setminus J})$ with $k'=0,...,K'$, where $K'\le C_3(R,I)| \BLam'|$. This gives:
$$\begin{array}{l}
\P\big(\dist\left[\Sigma\left(H_{\BLam}\right)\cap I,
\Sigma\left(H_{\BLam'}\right)\cap I\right]
\leq \eps|\Bmf_{\Z^d \setminus J} \big)\\
\le\sum_k^K\sum_{k'}^{K'} \P\big(|E^{(\BLam)}_{k}({\mathbf V}_J) -
E^{(\BLam')}_{k'}({\mathbf V}_{\Z^d \setminus J})|\le\eps\big|\Bmf_{\Z^d \setminus J} )
\\
\le\sum_k^K C_3(R,I)| \BLam'|\sup_{E\in I}
\P\big(|E^{(\BLam)}_{k}({\mathbf V}_J) - E|\le\eps\big|\Bmf_{\Z^d \setminus J} )
\\
=\sum_k^K C_3(R,I)| \BLam'|\sup_{E\in I} \P\big(E^{(\BLam)}_{k}({\mathbf V}_{J})\in
[E-\eps, E+\eps]|\Bmf_{\Z^d \setminus J} ) .
  \end{array}
$$
As in the proof of Theorem \ref{ThmOneVol}, the probability
$$
\P\big(E^{(\BLam)}_{k}(\om)\in [E-\eps, E+\eps]| {\Bmf}_{\Z^d \setminus J} )\le C_3(R,I)| \BLam|
|\Pi_1\BLam| \nu(\eps) .
$$

A similar argument works for case (II).  Put together this gives the desired estimate,
where the factor
$${\operatornamewithlimits{\max}\limits_{j=1,2}}\;
\max\;\Big[\big|\Pi_j\BLam\big|,\big|\Pi_j\BLam'\big|\Big]
\eqno(4.5)
$$
accounts for the different geometric possibilities in Lemma \ref{SepLem}.

\qed

\pmn
{\bf Acknowledgments.} The authors thank The Isaac Newton Institute, University of
Cambridge, for hospitality during the programme
``Mathematics and Physics of the Anderson Localisation:
50 years after'' (July--December, 2008). PS thanks Universit\'e Paris VII and the DFG for supporting travel and
YS thanks IHES, Bures-sur-Yvette, for hospitality during visits in 2008.

%
%

\end{document}